\documentclass[aps,prr,amssymb,reprint,twocolumn,superscriptaddressgroupaddress]{revtex4-1}
\usepackage{amsmath}
\usepackage{hyperref}
\usepackage[english]{babel}
\usepackage[utf8]{inputenc}
\usepackage{float}
\usepackage{graphicx}
\usepackage{braket}
\usepackage{xcolor}

\begin{document}
	\title{Time-reversal-invariant $C_2$-symmetric higher-order topological superconductors}
	\author{DinhDuy Vu}
	\author{Rui-Xing Zhang}
	\email{ruixing@umd.edu}
	\author{S. Das Sarma}
	\affiliation{Condensed Matter Theory Center and Joint Quantum Institute, Department of Physics, University of Maryland, College Park, Maryland 20742, USA}
	\begin{abstract}
		We propose a minimal lattice model for two-dimensional class DIII superconductors with $C_2$-protected higher-order topology. While this class of superconductors cannot be topologically characterized by symmetry eigenvalues at high symmetry momenta, we propose a simple Wannier-orbital-based real-space diagnosis to unambiguously capture the corresponding higher-order topology. We further identify and characterize a variety of conventional topological phases in our minimal model, including a weak topological superconductor and a nodal topological superconductor with chiral-symmetry protection. The disorder effect is also systematically studied to demonstrate the robustness of higher-order bulk-boundary correspondence. Our theory lays the groundwork for predicting and diagnosing $C_2$-protected higher-order topology in class DIII superconductors.
	\end{abstract}
	
	\maketitle
	
	\section{Introduction}
	
	Anyons are stable exotic quasiparticles with unconventional statistical braiding properties and serve as the cornerstone for topological quantum computation \cite{Sarma2005,Kitaev2006,Nayak2008,Alicea2011}. The most well-known anyonic quasiparticle is the zero-dimensional (0D) Majorana zero mode (MZM), which could in principle emerge as a vortex bound state of a two-dimensional (2D) $p+ip$ chiral topological superconductor (TSC) \cite{Read2000}, or as the end mode of a one-dimensional (1D) $p$-wave TSC \cite{Kitaev2001,Lutchyn2010}. Remarkable experimental progress has been made towards realizing Majorana physics in superconducting Rashba nanowires \cite{Mourik2012,Zhang2018} and recently in iron-based superconductors \cite{Wang2018b}, where promising signatures of MZMs such as zero-bias tunnel conductance peaks have been observed. However, a ``smoking-gun" measurement of confirming MZM's anyonic nature is missing. Therefore, the experimental existence of MZMs is still not conclusive, which calls for more efforts on topological Majorana physics, perhaps in different systems.  
	
	Recently, there has been growing interest in understanding higher-order topology \cite{Benalcazar2017,Schindler2018,Langbehn2017,Khalaf2018}, where anomalous boundary physics could show up on the $(d-n)$-dimensional boundary of a $d$-dimensional topological system with $1<n\le d$. In particular, a 2D higher-order TSC is defined to host MZMs on its 0D geometric corners ($d=n=2$ here) \cite{Wang2018a,Yan2018,Shapourian2018,Liu2018,Hsu2018,Pan2019,Zhang2019,Wu2019}, which is a novel promising Majorana platform. Such corner-localized MZMs are proposed to exist in iron-based superconductors \cite{Zhang2019,Wu2019,Wu2020} and doped WTe$_2$ \cite{Hsu2019}. However, in most of these proposals, the corner MZMs are relatively ``fragile" and can be easily removed upon closing the edge energy gap. The robustness of the corner MZMs, however, can be enhanced against symmetry-preserving edge perturbations if we can introduce additional crystalline-symmetry protections \cite{Khalaf2018,Wang2018a,Yan2018,Shapourian2018,Liu2018,Hsu2018,Pan2019,Zhang2019,Wu2019,Wu2020,Hsu2019}. Several groups have proposed various symmetry indicators to classify such symmetry-protected higher-order TSCs based on symmetry eigenvalues at high symmetry momenta \cite{Khalaf2018a,Ono2019,Geier2019,Shiozaki2019,Skurativska2020}. In this context, class DIII superconductors with two-fold rotational symmetry $C_2$ are special in the sense that they always host the same $C_2$ symmetry eigenvalues at high symmetry momenta for all possible topologically distinct phases. Consequently, this class of superconductors does not admit an indicator-based classification \cite{Ono2020}. While the absence of an indicator does not necessarily imply trivial topology, understanding and characterizing higher-order topology for this symmetry class remains an important open question in spite of considerable recent activities in the topological classification of superconductors.
	
	Recently, a real-space building scheme of 2D class D topological superconductors has been proposed \cite{Zhang2020}. The scheme is based on different stackings of 1D topological Kitaev chains which can lead to intrinsic Majorana corner modes. Interestingly, the concept of building a higher-order topological superconductor from Kitaev building blocks also allows a topology diagnosis based on the real-space distribution of Wannier orbitals and lattice sites -- the Majorana counting rule. Motivated by this novel idea, we aim to expand the counting rule to class DIII superconductors with spin degrees of freedom and time-reversal invariance. In particular, our real-space counting predicts that the presence of corner Majorana Kramers pairs (MKPs) only depends on the information of lattice geometry and Bogoliubov-de Gennes (BdG) Wannier orbitals, but not the microscopic details of Cooper pairings. For demonstration, we propose a minimal example of a 2D class DIII higher-order topological superconductor with $C_2$ symmetry protection, which is beyond the application of symmetry indicators. Despite its simplicity, our minimal model hosts surprisingly rich topological physics, including a higher-order TSC phase with 0D corner-localized Majorana Kramers pairs (MKPs), a weak TSC phase with 1D Majorana edge band, and a chiral-symmetry-protected nodal TSC phase with edge Majorana flat band. Noticeably, all topological phases in this model can be characterized with our class DIII counting rule. 
	
   The paper is organized as follows. In Sec. II, we introduce some key concepts and derive the Majorana counting rule for class DIII superconductors. In Sec. III, we introduce our minimal class DIII model with $C_2$-protected higher-order topology. We then map out the topological phase diagram for our minimal model, predict the topological nature of each phase (i.e. the higher-order TSC, the weak TSC, and the nodal TSC) with our counting rule, and further confirm our prediction numerically. The stability of the higher-order topological phase against disorder is studied, which directly proves the higher-order bulk-boundary correspondence. The conclusion is provided in Sec. IV. 
	
	\section{Majorana counting rule\\ in class DIII}
	
	\begin{figure}
	    \centering
	    \includegraphics[width=0.47\textwidth]{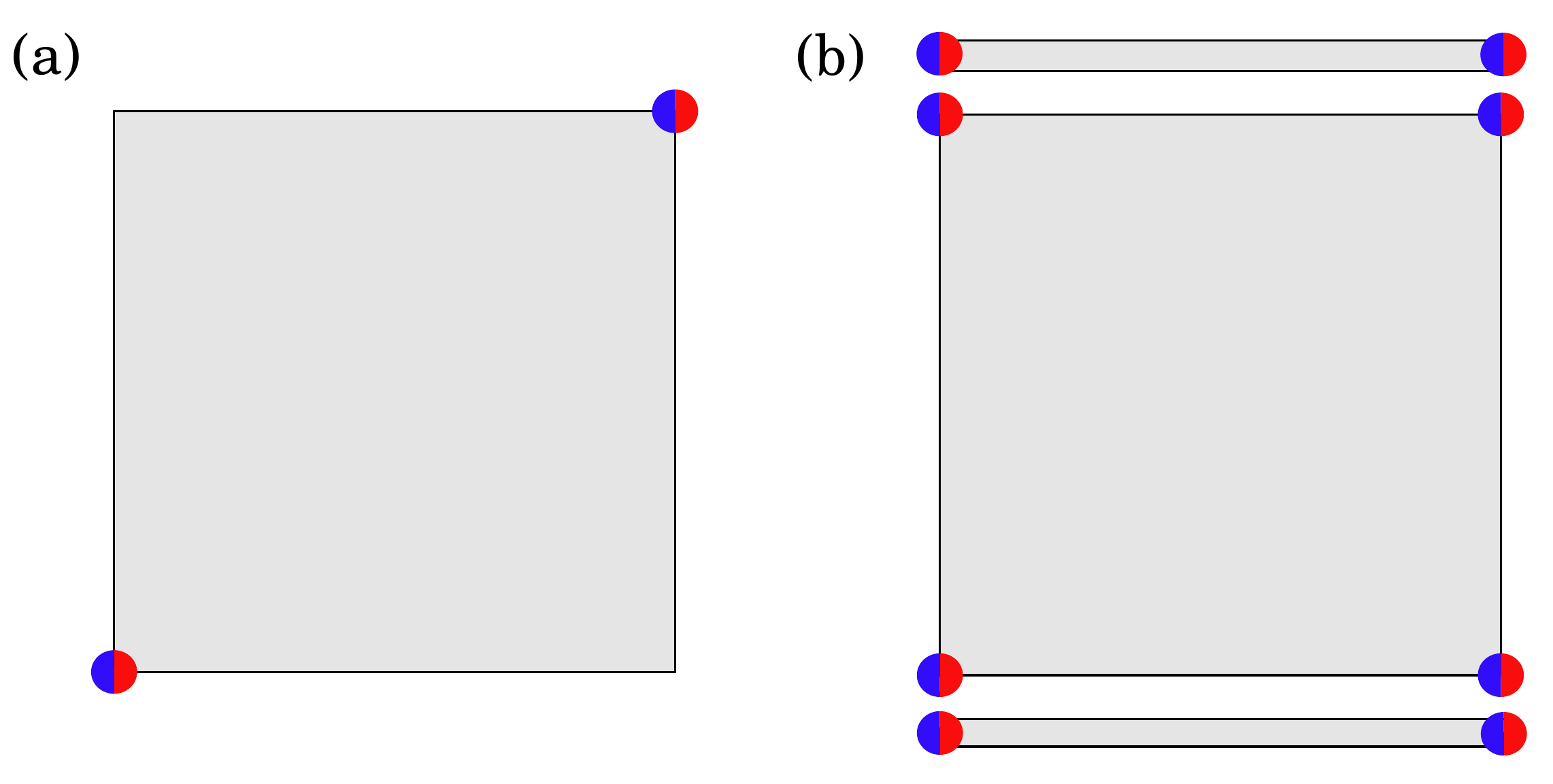}
	    \caption{(a) $C_2$-protected higher-order topology with two MKPs on two $C_2$-related corners. (b) A $C_2$-symmetric system with four corner MKPs is topological trivial, since the corner MKPs can be removed upon symmetrically attaching 1D class DIII TSCs on the boundary.}
	    \label{corner}
	\end{figure}
	
	We define a 2D class DIII superconductor to host $C_2$-protected second-order topology if it hosts two robust Kramers pairs of 0D Majorana bound states that are localized around two {\it $C_2$-related} lattice sites on the boundary. Most likely, the boundary MKPs will sit around certain sample corners, as shown in Fig.~\ref{corner}(a). It is easy to see that such boundary/corner MKPs cannot be eliminated by any perturbation that preserves both adiabaticity and $C_2$. In contrast, if MKPs appear at all four corners, such a $C_2$-invariant system is NOT bulk higher-order topological, since the corner MKPs can be eliminated symmetrically by attaching 1D class DIII TSCs on the boundary [see Fig.~\ref{corner}(b)]. Therefore, we can phenomenologically characterize the presence of higher-order topology by the number of MKPs ${\cal N}$ on the ``half boundary", which, for a rectangular geometry, can be chosen as a collection of one $x$-edge, one $y$-edge, and their shared corner. In particular, an odd ${\cal N}$ is the necessary condition for achieving $C_2$-protected higher-order topology.
	
	To calculate ${\cal N}$, it is convenient to map bulk electron and hole degrees of freedom into a pair of Majorana fermions $\alpha_{{\bf R},s}$ and $\beta_{{\bf R},s}$, where ${\bf R}$ is the real-space position and $s$ is the spin index. Note that the bulk Majorana operators necessarily come in Kramers pairs due to the time-reversal symmetry. This Majorana representation allows us to generalize the concept of ``Kitaev limit"\cite{Zhang2020} to the time-reversal-invariant version, which is achieved for a class DIII superconductor when every bulk MKP formed by electron and hole Kramers pairs is connected to {\it exactly} one other MKP via Majorana bonds. Notably, distinct from the situation in class D, the MKP's spin degrees of freedom enable two different types of Majorana bondings:
	\begin{itemize}
		\item equal-spin bonding: Majorana fermions with the same spin indices are bonded, which effectively leads to $p$-wave Cooper pairing;
		\item opposite-spin bonding: Majorana fermions with opposite spin indices are bonded, which effectively leads to $s$-wave/$d$-wave Cooper pairing.
	\end{itemize}   
	We prove in Appendix A, for every Kitaev limit, there exists a Kramers pair of maximally localized BdG Wannier orbitals sitting at the center of each bond, whose positions of Wannier centers does not rely on the explicit bonding type. We emphasize that a Kitaev limit is necessarily gapped in its bulk spectrum, since every bulk MKP is paired with one another. Nevertheless, unpaired zero-energy MKPs can always appear on the system boundary, when some Majorana bonds are cut by imposing the open boundary. Therefore, reducing a target system to its Kitaev limit greatly facilitates us to calculate ${\cal N}$ by just counting the unpaired boundary MKPs.
	
	We now proceed to derive the value of ${\cal N}$ for a general Kitaev limit following the approach in Ref. \cite{Zhang2020}. In a $C_2$-invariant unit cell, there exists four inequivalent maximal Wyckoff positions, denoted as $\mathbf{q}_{1a}=(0,0)$, $\mathbf{q}_{1b}=(\frac{1}{2},0)$, $\mathbf{q}_{1c}=(\frac{1}{2},\frac{1}{2})$ and $\mathbf{q}_{1d}=(0,\frac{1}{2})$. If a Kramers pair of BdG Wannier orbital pair sits at a generic Wyckoff position ${\bf q} \neq {\bf q}_{1i}$ (for $i=a,b,c,d$), $C_2$ symmetry will require another Kramers pair to occur at the Wyckoff position ${\bf q}'=C_2{\bf q}$. Then these two Wannier orbital pairs can always be symmetrically moved to any of the four maximal Wyckoff positions. As a result, only Wannier orbitals at the maximal Wyckoff positions will play a role in our discussion.

	We note that ${\cal N}$ is contributed by both $n_i^{M}$ and $n_i^{W}$, which count the numbers of MKPs and Wannier orbital pairs at a maximal Wyckoff position $\mathbf{q}_i$ inside one unit cell, respectively. In particular, Wannier orbitals placed at ${\bf q}_{1i}$ with $i=b,c,d$ will necessarily
    lead to boundary ``dangling" ones in the presence of an open geometry. Such a dangling Wannier orbital pair indicates missing Majorana bonds on the boundary, which thus corresponds to one unpaired MKP. For a finite system with $L_x\times L_y$ unit cells, it is straightforward to show that $n_i^W$ indicates (i) $0$ unpaired MKP for $i=a$; (ii) $n_b^WL_y$ unpaired MKPs for $i=b$; (iii) $n_c^W(L_x+L_y-1)$ unpaired MKPs for $i=c$; (iv) $n_d^W L_x$ unpaired MKPs for $i=d$ on the half boundary.
	
	On the other hand, we will also need to remove some ``dangling" lattice sites on the boundary, as well as the corresponding MKPs sitting on top, to respect $C_2$. This counting procedure is similar to that for the dangling Wannier orbitals, which leads us to the removal of (i) $0$ MKP for $i=a$; (ii) $n_b^ML_y$ MKPs for $i=b$; (iii) $n_c^M(L_x+L_y-1)$ MKPs for $i=c$; (iv) $n_d^ML_x$ MKPs for $i=d$ on the half boundary. Taking into account both the gain and loss of boundary MKPs, we find that
	\begin{equation}\label{eq2}
	\begin{split}
	\mathcal{N} = L_x\Delta_{d}+L_y\Delta_{b} +(L_x+L_y-1)\Delta_c,
	\end{split}
	\end{equation}
	where we have defined a set of Majorana counting numbers labeled by the Wyckoff position index $i=a,b,c,d$,
	\begin{equation}
	\Delta_i=n_i^W-n_i^M.
	\end{equation}
	
    For our purpose, we hope to rule out weak topological phenomena, which can be achieved by imposing $\Delta_{b}+\Delta_{c}\equiv\Delta_{d}+\Delta_{c}\equiv0$~(mod 2) \cite{Zhang2020}. Combining with our requirement ${\cal N} \equiv 1$ (mod 2), we obtain a compact Majorana counting rule to guarantee the $C_2$-protected higher-order topology:
    \begin{equation}\label{eq3}
    	\Delta_{b,c,d} \equiv 1~\text{(mod 2)}.
    \end{equation}
   When the counting rule in Eq.~\eqref{eq3} is satisfied, we expect the system to host two zero-energy MKPs at two $C_2$-related corners. On the other hand, if $\Delta_{d}+\Delta_{c}$ and/or $\Delta_{b}+\Delta_{c}$ are odd, the system has weak topology characterized by flat-band-like Majorana edge modes protected by certain lattice translation symmetry. Trivial higher-order/weak topology is indicated by $\Delta_{b,c,d} \equiv 0$~(mod 2). Finally, a system that fails to admit a Kitaev limit is expected to be gapless in the bulk spectrum.

   We emphasize that while this Majorana counting rule is derived in the Kitaev limit, we expect it to hold for a general Wannierizable superconductor. The procedure for reducing a non-Kitaev superconductor to its Kitaev limit generally involves smooth deformations such as minimizing the localization length of Wannier functions and adiabatic movements of Wannier orbitals. Clearly, our Majorana counting rule is topologically immune to these deformations as long as they are adiabatic and symmetric.

	\begin{figure}
		\includegraphics[width=0.45\textwidth]{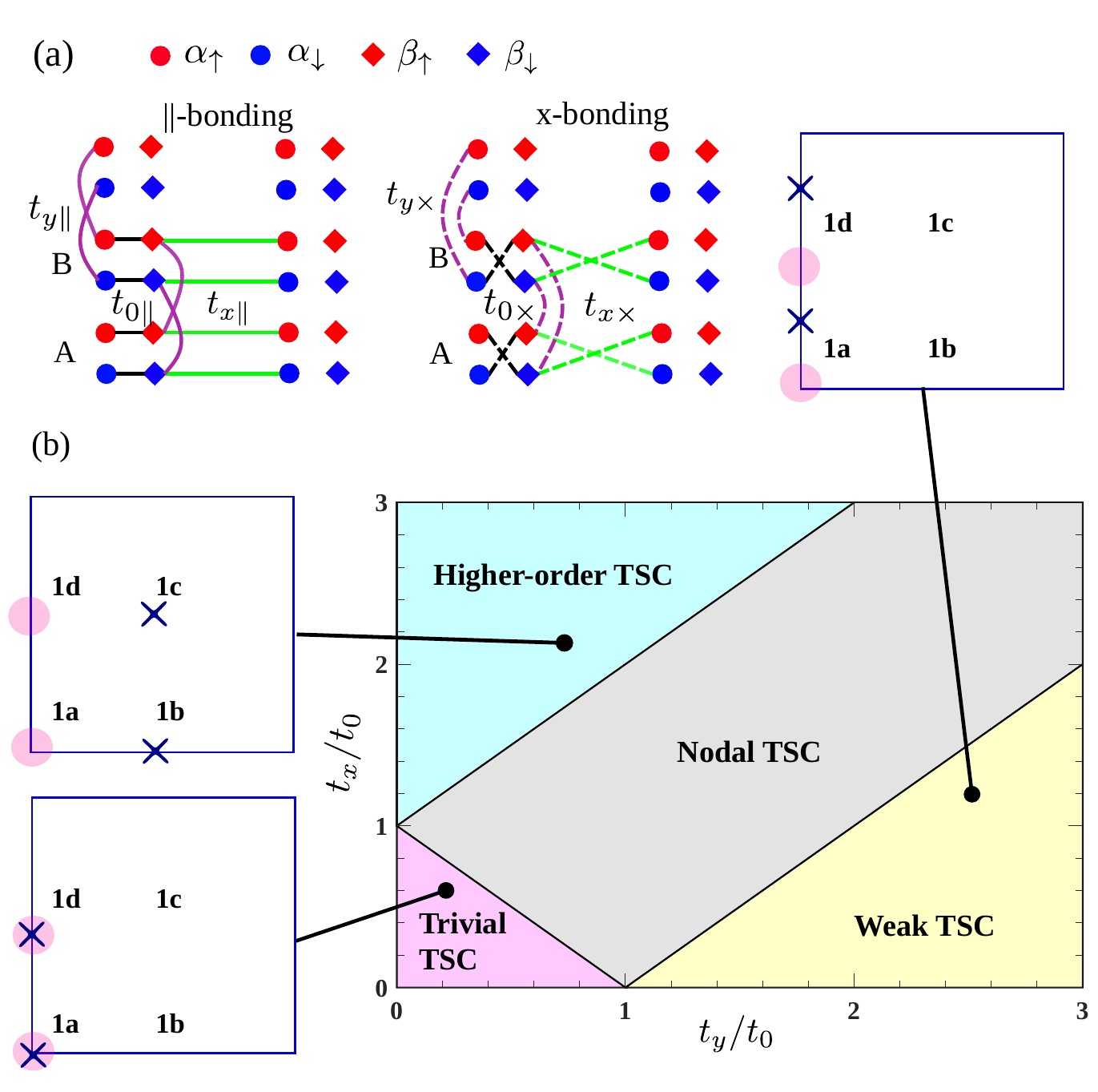}
		\caption{(a) Six-parameter bonding scheme of the model Hamiltonian. Each dot/diamond is a Majorana and each line is an equal-spin ($\parallel$) or opposite-spin ($\times$) bonding. (b) Phase diagram based on the effective bonding strength. Each gapped phase is classified according to the corresponding Kitaev limit with positions of the MKPs (dots) and Wannier orbital pairs (crosses) depicted in a unit cell.}\label{fig1}
	\end{figure}	
	
   \section{Time-reversal-invariant $C_2$-symmetric model}	
	
	We now introduce our minimal model to demonstrate the real-space diagnosis for $C_2$-protected time-reversal-invariant higher-order topology,
    \begin{equation}
		\begin{split}
		H=i\sum_{\mathbf{R},l}&\left[ \beta^l_{\mathbf{R},s}\left(t_{0\parallel}^l\sigma_0+it_{0\times}^l\sigma_y\right)_{ss'}\alpha^l_{\mathbf{R},s'}\right.\\& +\left. \beta^l_{\mathbf{R},s}\left(t_{x\parallel}^l\sigma_0+it_{x\times}^l\sigma_y\right)_{ss'} \alpha^l_{\mathbf{R}+\mathbf{a},s'}\right]\\
		+i\sum_{\mathbf{R}} & \left[ \beta_{\mathbf{R},s}^A\left(t_{y\parallel}\sigma_z+t_{y\times}\sigma_x\right)_{ss'}\beta_{\mathbf{R},s'}^B \right.\\& \left.- \alpha_{\mathbf{R},s}^B\left( t_{y\parallel}\sigma_z-t_{y\times}\sigma_x\right)\alpha_{\mathbf{R}+\mathbf{b},s'}^A \right];
		\end{split}\label{Hamiltonian}
	\end{equation}
	where  $l=A/B$ is the sub-lattice index, $\mathbf{a}$ and $\mathbf{b}$ are lattice constants along $x$ and $y$ directions, and $\sigma_{i}$ ($i=0,x,y,z$) are the Pauli matrices for spins. The subscript $\parallel$ ($\times$) denotes the equal (opposite)-spin bonding. The Hamiltonian can be transformed to fermion operators using $\alpha_{s,\mathbf{R}}^l = (c^{l \dagger}_{s,\mathbf{R}}+c^l_{s,\mathbf{R}})/\sqrt{2}$ and  $\beta^l_{s,\mathbf{R}} = i(c^{l \dagger}_{s,\mathbf{R}}-c^l_{s,\mathbf{R}})/\sqrt{2}$. Without loss of generality, we choose $t_{0\parallel}^A=t_{0\parallel}^B=t_{0\parallel}$, $t_{0\times}^A=-t_{0\times}^B=t_{0\times}$, $t_{x\parallel}^A=t_{x\parallel}^B=t_{x\parallel}$, and $t_{x\times}^A=-t_{x\times}^B=-t_{x\times}$ for convenience. The bonding scheme of Eq.~\ref{Hamiltonian} within a unit cell is depicted in Fig.~\ref{fig1}(a). In the fermion representation, the momentum-space Bogoliubov-de Genes Hamiltonian is
	\begin{align}
	& \mathcal{H}(\mathbf{k})= \left[-(t_{0\parallel}+t_{x\parallel}\cos k_x)\tau_z + t_{x\parallel}\sin k_x \tau _y\right]\sigma_0\gamma_0 \nonumber \\
	&+ \left[(t_{0\times}+t_{x\times}\cos k_x)\tau_y + t_{x\times}\sin k_x \tau _z\right]\sigma_y\gamma_z\\
	& +(t_{y\parallel}\tau_x\sigma_z-t_{y\times}\tau_0\sigma_x)\left[(1-\cos k_y)\gamma_y + \sin k_y \gamma_x\right]/2 \nonumber \\
	&+(t_{y\parallel}\tau_0\sigma_z-t_{y\times}\tau_x\sigma_x)\left[-(1+\cos k_y)\gamma_y + \sin k_y \gamma_x\right]/2, \nonumber
	\end{align} 
	where $\tau$ and $\gamma$ are Pauli matrices representing the particle-hole and sub-lattice degrees of freedom. It is easy to check that $\mathcal{H}(\mathbf{k})$ is invariant under time-reversal operator $\Theta=i\sigma_y\mathcal{K}$, particle-hole operator  $\mathcal{P} = \tau_x \mathcal{K}$, and rotation $C_2=i\tau_z\sigma_z\left[e^{ik_y}(\gamma_z+\gamma_0)+(\gamma_0-\gamma_z) \right]/2$. Here $\mathcal{K}$ is the complex conjugation.
	
	The gap closing condition of $\mathcal{H}(\mathbf{k})$ is determined by three independent parameters: $t_0=(t_{0\times}^2+t_{0\parallel}^2)^{1/2}$, $t_x=(t_{x\times}^2+t_{x\parallel}^2)^{1/2}$, $t_y=(t_{y\times}^2+t_{y\parallel}^2)^{1/2}$ (see Appendix B). Physically, $t_0$ characterizes the net bonding strength between onsite MKPs, while $t_x$ and $t_y$ characterize the bonding strength for nearest-neighbor MKPs along $x$ and $y$ directions, respectively. The Wannier orbital sits at the center of the strongest Majorana bond when other bonds can be adiabatically reduced to zero. For example, if $t_x>t_{0,y}$, we can immediately tell the existence of Wannier orbital pairs sitting at both $1c$ and $1d$. We note that the position of Wannier orbitals is independent of the spin texture of the bondings.
	
	Meanwhile, we are able to analytically map out the phase diagram in terms of $t_x/t_0$ and $t_y/t_0$, as shown in Fig.~\ref{fig1}(b). Below, we ascertain the topological nature of each phase by smoothly deforming into the corresponding Kitaev limit and applying our Majorana counting rule to perform the topological diagnosis. While such a Kitaev reduction is not necessary for the topological diagnosis, this procedure makes it easier to directly ``read out" the position information of Wannier orbitals, greatly facilitating the application of the counting rule.
	
	\subsection{Higher-order TSC phase}
	
	\begin{figure}
	\includegraphics[width=0.5\textwidth]{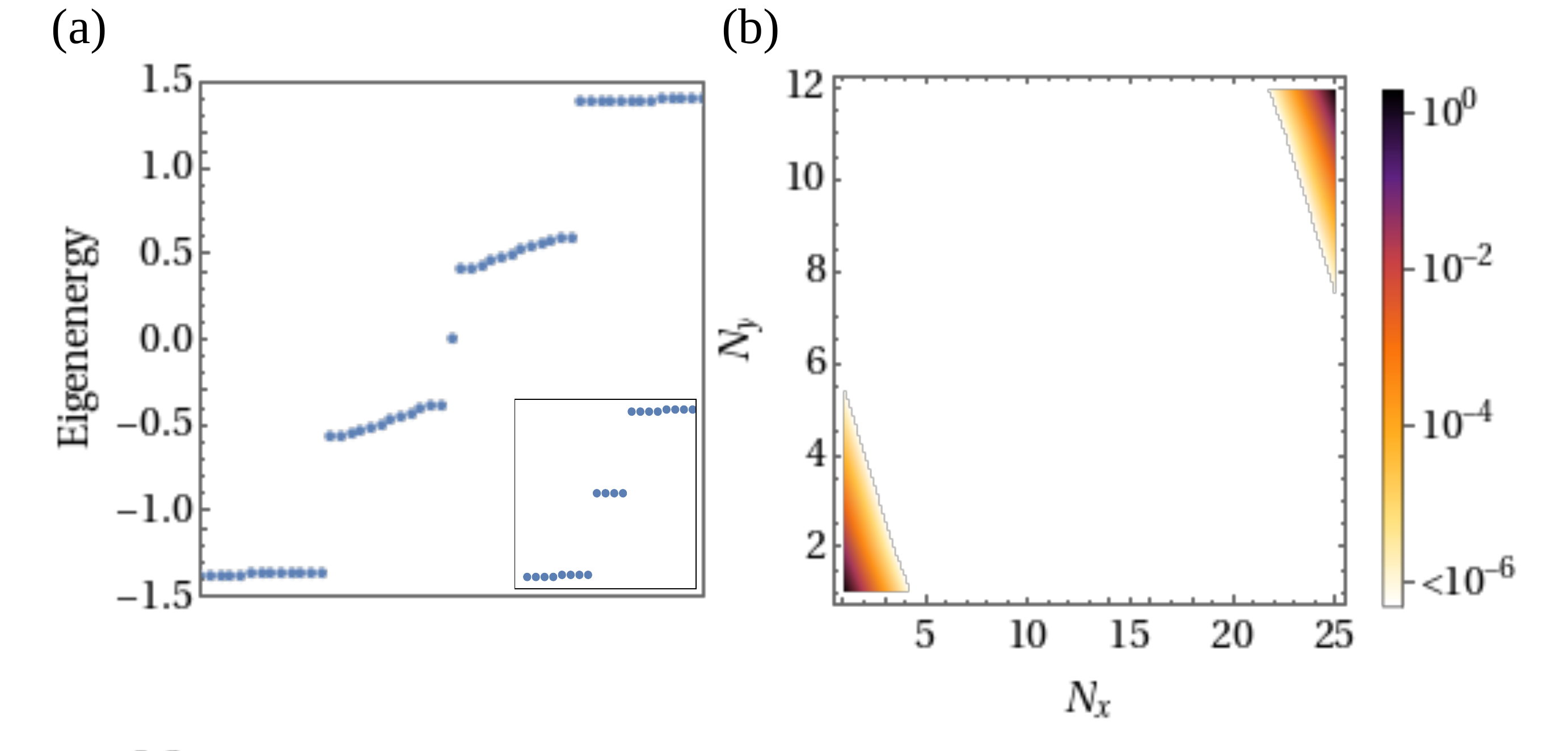}
	\caption{(a) Energy spectrum in a finite array of $25\times25$ atoms in the higher-order TSC phase. $N_x$ and $N_y$ are unit-cell indices along the $x$ and $y-$directions, $t_0=0.2$, $t_x=2$, $t_y=0.5$ and $t'=0.1$. The inset zooms in the four zero-energy states. (b) Spatial profile of the zero-energy modes of the same system.}\label{fig2}
    \end{figure}	

    \begin{figure}
	\includegraphics[width=0.42\textwidth]{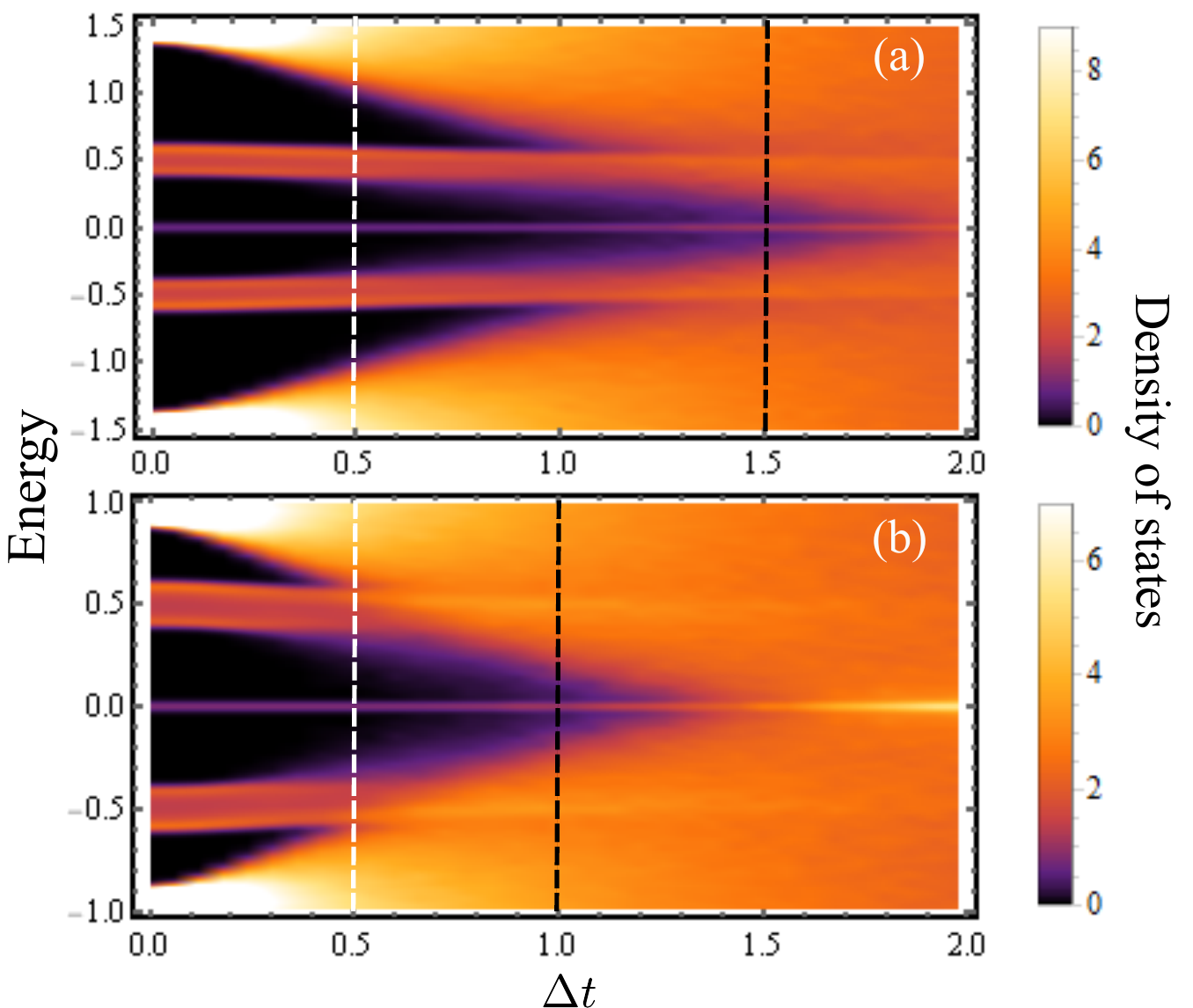}
	\caption{Density of states with respect to the disorder strength of a $25\times25$-unit-cell finite system. (a) $t_x=2, t_y=0.5, t'=0.1$. (b) $t_x=1.5,t_y=0.5,t'=0.1$. The value of the bulk mid-gap ($t_x-t_y$) is marked by the black dashed lines, which is also the critical disorder strength that destroys the corner MKPs; while the edge mid-gap $t_y$ is marked by the white dashed lines.}\label{fig3}
    \end{figure}	
	
	In the regime $t_x>t_0+t_y$, we deform the Hamiltonian to the Kitaev limit where only $t_x>0$ and $t_0, t_y\to \epsilon$ with $\epsilon \ll t_x$. As shown in Fig. \ref{fig1}(b), while the two MKPs within one unit cell both originate from maximal Wyckoff positions $1a$ and $1d$, the Wannier orbital pairs sit at $1b$ and $1c$. Therefore, the counting numbers are $\Delta_{d}=-1$ and $\Delta_{b,c}=1$, and thus our Majorana counting rule predicts this gapped phase as a $C_2$-protected higher-order TSC. 
	
	To confirm the above predictions, we now simulate this phase numerically in an open-boundary finite-size system. Without loss of generality, we choose $t_x=t_{x\times}$ and $t_y=t_{y\times}$ to be opposite-spin Majorana bonding ($s$-wave-like pairing) and $t_0=t_{0\parallel}$ to act as the chemical potential. For a more realistic manifestation of the corner modes, we introduce a small perturbation along the vertical direction.
	\begin{equation}\label{eq7}
	H'=it'\sum_{\mathbf{R}} \alpha_{\mathbf{R},s}^A\left(\sigma_x\right)_{ss'}\alpha_{\mathbf{R},s'}^B + \beta_{\mathbf{R},s}^B\left(\sigma_x\right)_{ss'}\beta_{\mathbf{R}+\mathbf{b},s'}^A,
	\end{equation}
	which should slightly delocalize possible topological bound states in our system.
	
	As shown in Fig.~\ref{fig2}(a), the energy spectrum on a $25\times 25$ lattice shows 4 degenerate Majorana modes at zero energy. By examining their spatial profiles in Fig.~\ref{fig2}(b), the four Majorana modes are divided into two spatially separated MKPs which are localized exponentially around the two $C_2$-related geometric corners. This unambiguously establishes the $C_2$-protected higher-order topology in this phase, confirming the prediction from the Majorana counting rule.
	
	On the other hand, this higher-order TSC phase manifests itself as an example that goes beyond the symmetry-eigenvalue-based topological characterization scheme. This is simply because a $C_2$-invariant momentum is also a time-reversal-invariant momentum in momentum space. For class DIII systems, each Kramers pair at every high symmetry momentum carries $(+i,-i)$ as their $C_2$ eigenvalues. Therefore, the symmetry information for such systems is essentially ``featureless" and remains the same across possible topological phase transitions. Indeed, a rather complete symmetry indicator theory for BdG systems in all space groups has recently concluded the absence of momentum-space indicators for this symmetry class \cite{Ono2020}. By contrast, our Majorana counting rule is based on a real-space description, thus fundamentally avoiding the difficulty caused by the featureless symmetry data. 

	\begin{figure*}[ht!]
	\includegraphics[width=0.93\textwidth]{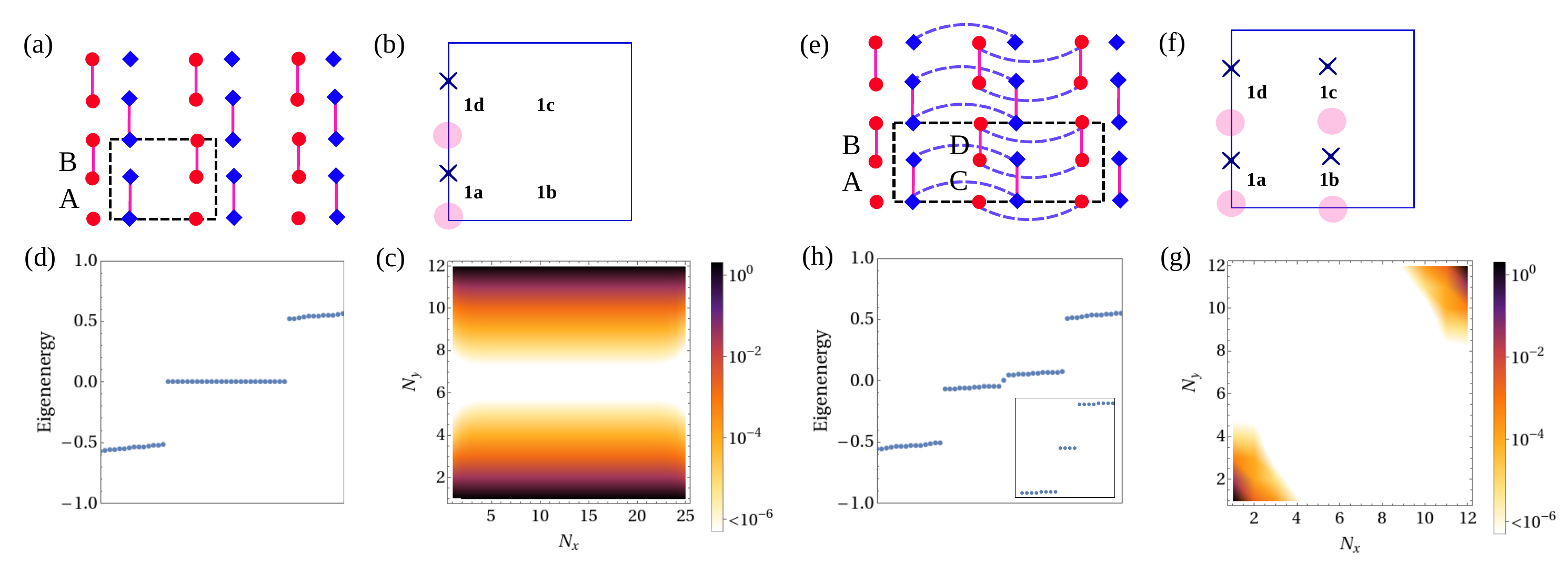}
	\caption{(a)-(d): The Kitaev limit where each dot/diamond represents a Majorana Krammer pair and each line represents a $2\times2$ bonding tensor, the Wannier orbital spatial configuration, the energy spectrum, and the spatial density of zero-energy modes in the weak TSC phase with $t_0=0.2$, $t_x=0.5$, $t_y=1$ and $t'=0.1$. (e)-(h): Same as (a)-(d) but in the higher-order TSC with an additional dimer bond along $x$-direction $t'_x=0.1$. The inset in (g) shows a zoom-in plot of the corner MKPs. The unit cell in (h) is doubled along $x$ so the range of $N_x$ is divided by two.\label{dimmer}}
   \end{figure*}    

	We now study the robustness of the higher-order TSC phase against local disorder. First, we note that there are two important energy scales in our system in the $t_0\rightarrow 0$ limit: (i) the bulk gap $2|t_x-t_y|$ and (ii) the edge gap $2|t_y|$. From the higher-order bulk-boundary correspondence, we expect that the corner MKPs could be  destroyed {\it only} when the disorder strength exceeds the bulk energy gap, if the disorder globally preserves the protecting symmetries.

    Numerically, we introduce chemical potential fluctuations by randomizing the strength of on-site bonding $t_0$ following a Gaussian ensemble, which is centered at zero and with a standard deviation of $\Delta t$. For each value of $\Delta t$, we average the density of states over 100 random configurations. We now consider two different sets of bonding parameters such that the two corresponding systems share the same edge gaps but differ in their bulk gaps. As shown in Figs. \ref{fig3}(a) and (b), the critical disorder strength that suppresses corner Majorana modes is close to the value of the bulk gap (black dashed lines), and not the edge gap (white dashed lines). This clearly demonstrates the bulk origin and the consequent stability of higher-order topology in our system.

	\subsection{Weak TSC phase}
	In the regime $t_y>t_x+t_0$, we adiabatically approach the Kitaev limit by sending $t_x,t_0\to \epsilon$ with $\epsilon \ll t_y$. In this limit, the system resembles a set of decoupled $y-$directional 1D time-reversal-invariant TSCs \cite{Zhang2013} stacking along $x$ direction [see Fig.~\ref{dimmer}(a)], manifesting itself as a weak topological superconductor protected by the $x-$directional translational symmetry. We immediately see that $\Delta_{d}=-1$ while $\Delta_{b,c}=0$ [see Fig.~\ref{dimmer}(b)], indicating the absence of higher-order topology. The presence of weak topology is also indicated by the oddness of $\Delta_d+\Delta_c$. As each 1D TSC hosts two end MKPs, the collection of edge MKPs in our weak TSC phase forms two pairs of time-reversal-invariant Majorana flat bands that are individually localized at the upper and lower edges. To see this, we numerically plot the energy spectrum of a $25\times25$-atom lattice in Fig.~\ref{dimmer}(c). Specially, each upper and bottom edge hosts $2L_x$ localized zero-energy states, which confirms the existence of the Majorana flat bands in Fig.~\ref{dimmer}(d). 
	
	The weak topology is spoiled by breaking the translational symmetry along $x$ direction. This can be achieved with the following dimer perturbation, 
	\begin{equation}
	H'_x=it'_x\sum_{m,n,\sigma}(\beta_{m,n}^{A/B,\sigma}\beta_{m,n}^{C/D,-\sigma} + \alpha_{m,n}^{C/D,\sigma}\alpha_{m+1,n}^{A/B,-\sigma}),
	\end{equation}
	which doubles the unit cell along $x$-direction, as shown in Fig.~\ref{dimmer}(e). Interestingly, such unit-cell doubling directly leads to an occupation of one MKP and zero Wannier orbital pair for all four maximal Wyckoff positions. Now we have $\Delta_{b,c,d}=-1$ [see Fig.~\ref{dimmer}(f)] and thus the higher-order topological condition is satisfied. In Figs.~\ref{dimmer}(g) and (h), we numerically calculate the energy spectrum with $H'_x$ and indeed find two $C_2$-related corner-localized MKPs that signals the higher-order topology, agreeing with our counting rule prediction. This is a demonstration of the versatility of our extended counting rule that can be used as an efficient real-space diagnosis and a powerful tool for model construction.
	
	\subsection{Nodal TSC phase}
	
\begin{figure}
		\includegraphics[width=0.5\textwidth]{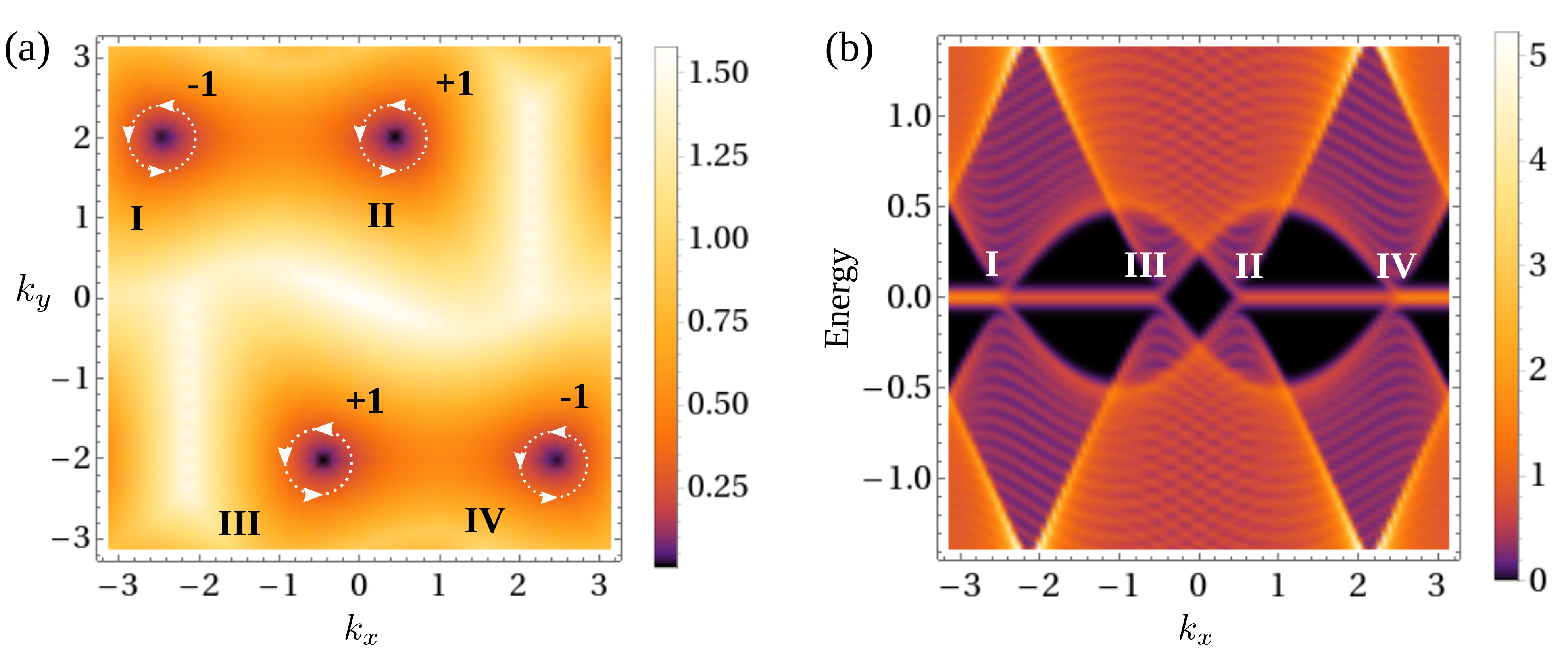}
		\caption{Nodal SC at $t_x=t_0=1,t_y=1.5$ and $\theta_0=\theta_y=1,\theta_x=2$. (a) Bulk-gap across the Brillouin zone having four Dirac points with non-trivial winding numbers. (b) Energy spectrum of the same system in the ribbon configuration. The Majorana flat bands arise from the non-trivial topology of Dirac points and are protected by the chiral symmetry.}\label{fig4}
	\end{figure}

    When $|t_x-t_0|\le t_y \le t_x+t_0$, we cannot send both $t_0$ and $t_y$ to zero, as shown in the phase diagram in Fig. \ref{fig1}(b). Then each bulk MKP is necessarily attached with more than one Majorana bonds. {\it This incapability of reducing to a Kitaev limit indicates a gapless nodal phase.} Indeed, we find the existence of four 2D Dirac nodes in this parameter regime, whose positions can be analytically solved as 
	\begin{equation}
	\begin{cases}
	k_y= 2\theta_y, k_x=\pm k_0 + (\theta_0-\theta_x)\\
	k_y=-2\theta_y, k_x=\pm k_0 + (\theta_x-\theta_0),
	\end{cases},
	\end{equation}
	where $k_0=\cos^{-1}\left[(t_y^2-t_0^2-t_x^2)/(2t_0t_x)\right]$, $\theta_0=\text{Arg}(t_{0\parallel}+it_{0\times})$, $\theta_x=\text{Arg}(t_{x\parallel}+it_{x\times})$ and $\theta_y=\text{Arg}(t_{y\parallel}+it_{y\times})$. For class DIII superconductors, the chiral symmetry ${\cal C}$, a combination of time-reversal and particle-hole symmetry, allows us to define a chiral winding number $\nu\in\mathbb{Z}$ \cite{Schnyder2011,Wong2013} to characterize the topological nature of the Dirac point (see Appendix C). Therefore, we numerically calculate the winding number, and find it to be non-zero for all four Dirac points. As shown in Fig.~\ref{fig4}(a), Dirac points I and IV share $\nu=-1$, while Dirac points II and III have $\nu=1$, which is compatible with the time-reversal symmetry.
	
	A nontrivial winding number necessarily implies the existence of edge Majorana flat band at zero energy \cite{Schnyder2011,Wong2013}. In Fig.~\ref{fig4}(b), we numerically calculate the energy spectrum for the nodal SC phase in a ribbon geometry. As expected from the bulk-boundary correspondence, edge Majorana flat bands lying at zero energy are found to connect Dirac points with opposite $\nu$, e.g. (I) with (II), and (III) with (IV). The existence of both bulk nontrivial winding number and robust edge Majorana flat bands together establish this gapless phase as a chiral-symmetry-protected nodal TSC phase.
	
	\section{Conclusion}
	We propose a minimal model Hamiltonian for higher-order topology in class DIII systems, which contains rather rich topological phenomena such as higher-order TSC phase, weak TSC phase, and nodal TSC phase. While a momentum-space characterization becomes inapplicable, we succeed in deciphering the higher-order topology in real space with our Majorana counting rule, by identifying the spatial information of BdG Wannier orbitals. Our proposed counting rule should serve as an important diagnostic principle in the search for higher-order-topology-based Majorana platforms.
	
	{\it Acknowledgement} - R.-X. Z. thanks Sheng-Jie Huang for helpful discussions. This work is supported by the Laboratory for Physical Sciences and Microsoft Corporation. R.-X. Z. is supported by a JQI Postdoctoral Fellowship.
	
	\bibliographystyle{apsrev4-1}
    \bibliography{Higher_order_TSC}		

\begin{thebibliography}{34}%
\makeatletter
\providecommand \@ifxundefined [1]{%
 \@ifx{#1\undefined}
}%
\providecommand \@ifnum [1]{%
 \ifnum #1\expandafter \@firstoftwo
 \else \expandafter \@secondoftwo
 \fi
}%
\providecommand \@ifx [1]{%
 \ifx #1\expandafter \@firstoftwo
 \else \expandafter \@secondoftwo
 \fi
}%
\providecommand \natexlab [1]{#1}%
\providecommand \enquote  [1]{``#1''}%
\providecommand \bibnamefont  [1]{#1}%
\providecommand \bibfnamefont [1]{#1}%
\providecommand \citenamefont [1]{#1}%
\providecommand \href@noop [0]{\@secondoftwo}%
\providecommand \href [0]{\begingroup \@sanitize@url \@href}%
\providecommand \@href[1]{\@@startlink{#1}\@@href}%
\providecommand \@@href[1]{\endgroup#1\@@endlink}%
\providecommand \@sanitize@url [0]{\catcode `\\12\catcode `\$12\catcode
  `\&12\catcode `\#12\catcode `\^12\catcode `\_12\catcode `\%12\relax}%
\providecommand \@@startlink[1]{}%
\providecommand \@@endlink[0]{}%
\providecommand \url  [0]{\begingroup\@sanitize@url \@url }%
\providecommand \@url [1]{\endgroup\@href {#1}{\urlprefix }}%
\providecommand \urlprefix  [0]{URL }%
\providecommand \Eprint [0]{\href }%
\providecommand \doibase [0]{http://dx.doi.org/}%
\providecommand \selectlanguage [0]{\@gobble}%
\providecommand \bibinfo  [0]{\@secondoftwo}%
\providecommand \bibfield  [0]{\@secondoftwo}%
\providecommand \translation [1]{[#1]}%
\providecommand \BibitemOpen [0]{}%
\providecommand \bibitemStop [0]{}%
\providecommand \bibitemNoStop [0]{.\EOS\space}%
\providecommand \EOS [0]{\spacefactor3000\relax}%
\providecommand \BibitemShut  [1]{\csname bibitem#1\endcsname}%
\let\auto@bib@innerbib\@empty
\bibitem [{\citenamefont {{Das Sarma}}\ \emph {et~al.}(2005)\citenamefont {{Das
  Sarma}}, \citenamefont {Freedman},\ and\ \citenamefont {Nayak}}]{Sarma2005}%
  \BibitemOpen
  \bibfield  {author} {\bibinfo {author} {\bibfnamefont {S.}~\bibnamefont {{Das
  Sarma}}}, \bibinfo {author} {\bibfnamefont {M.}~\bibnamefont {Freedman}}, \
  and\ \bibinfo {author} {\bibfnamefont {C.}~\bibnamefont {Nayak}},\ }\href
  {\doibase 10.1103/PhysRevLett.94.166802} {\bibfield  {journal} {\bibinfo
  {journal} {Phys. Rev. Lett.}\ }\textbf {\bibinfo {volume} {94}},\ \bibinfo
  {pages} {166802} (\bibinfo {year} {2005})}\BibitemShut {NoStop}%
\bibitem [{\citenamefont {Kitaev}(2006)}]{Kitaev2006}%
  \BibitemOpen
  \bibfield  {author} {\bibinfo {author} {\bibfnamefont {A.}~\bibnamefont
  {Kitaev}},\ }\href {\doibase 10.1016/j.aop.2005.10.005} {\bibfield  {journal}
  {\bibinfo  {journal} {Ann. Phys. (N. Y).}\ }\textbf {\bibinfo {volume}
  {321}},\ \bibinfo {pages} {2} (\bibinfo {year} {2006})}\BibitemShut {NoStop}%
\bibitem [{\citenamefont {Nayak}\ \emph {et~al.}(2008)\citenamefont {Nayak},
  \citenamefont {Simon}, \citenamefont {Stern}, \citenamefont {Freedman},\ and\
  \citenamefont {{Das Sarma}}}]{Nayak2008}%
  \BibitemOpen
  \bibfield  {author} {\bibinfo {author} {\bibfnamefont {C.}~\bibnamefont
  {Nayak}}, \bibinfo {author} {\bibfnamefont {S.~H.}\ \bibnamefont {Simon}},
  \bibinfo {author} {\bibfnamefont {A.}~\bibnamefont {Stern}}, \bibinfo
  {author} {\bibfnamefont {M.}~\bibnamefont {Freedman}}, \ and\ \bibinfo
  {author} {\bibfnamefont {S.}~\bibnamefont {{Das Sarma}}},\ }\href {\doibase
  10.1103/RevModPhys.80.1083} {\bibfield  {journal} {\bibinfo  {journal} {Rev.
  Mod. Phys.}\ }\textbf {\bibinfo {volume} {80}},\ \bibinfo {pages} {1083}
  (\bibinfo {year} {2008})}\BibitemShut {NoStop}%
\bibitem [{\citenamefont {Alicea}\ \emph {et~al.}(2011)\citenamefont {Alicea},
  \citenamefont {Oreg}, \citenamefont {Refael}, \citenamefont {{Von Oppen}},\
  and\ \citenamefont {Fisher}}]{Alicea2011}%
  \BibitemOpen
  \bibfield  {author} {\bibinfo {author} {\bibfnamefont {J.}~\bibnamefont
  {Alicea}}, \bibinfo {author} {\bibfnamefont {Y.}~\bibnamefont {Oreg}},
  \bibinfo {author} {\bibfnamefont {G.}~\bibnamefont {Refael}}, \bibinfo
  {author} {\bibfnamefont {F.}~\bibnamefont {{Von Oppen}}}, \ and\ \bibinfo
  {author} {\bibfnamefont {M.~P.}\ \bibnamefont {Fisher}},\ }\href {\doibase
  10.1038/nphys1915} {\bibfield  {journal} {\bibinfo  {journal} {Nat. Phys.}\
  }\textbf {\bibinfo {volume} {7}},\ \bibinfo {pages} {412} (\bibinfo {year}
  {2011})}\BibitemShut {NoStop}%
\bibitem [{\citenamefont {Read}\ and\ \citenamefont {Green}(2000)}]{Read2000}%
  \BibitemOpen
  \bibfield  {author} {\bibinfo {author} {\bibfnamefont {N.}~\bibnamefont
  {Read}}\ and\ \bibinfo {author} {\bibfnamefont {D.}~\bibnamefont {Green}},\
  }\href {\doibase 10.1103/PhysRevB.61.10267} {\bibfield  {journal} {\bibinfo
  {journal} {Phys. Rev. B - Condens. Matter Mater. Phys.}\ }\textbf {\bibinfo
  {volume} {61}},\ \bibinfo {pages} {10267} (\bibinfo {year}
  {2000})}\BibitemShut {NoStop}%
\bibitem [{\citenamefont {Kitaev}(2001)}]{Kitaev2001}%
  \BibitemOpen
  \bibfield  {author} {\bibinfo {author} {\bibfnamefont {A.~Y.}\ \bibnamefont
  {Kitaev}},\ }\href {\doibase 10.1070/1063-7869/44/10S/S29} {\bibfield
  {journal} {\bibinfo  {journal} {Physics-Uspekhi}\ }\textbf {\bibinfo {volume}
  {44}},\ \bibinfo {pages} {131} (\bibinfo {year} {2001})}\BibitemShut
  {NoStop}%
\bibitem [{\citenamefont {Lutchyn}\ \emph {et~al.}(2010)\citenamefont
  {Lutchyn}, \citenamefont {Sau},\ and\ \citenamefont {{Das
  Sarma}}}]{Lutchyn2010}%
  \BibitemOpen
  \bibfield  {author} {\bibinfo {author} {\bibfnamefont {R.~M.}\ \bibnamefont
  {Lutchyn}}, \bibinfo {author} {\bibfnamefont {J.~D.}\ \bibnamefont {Sau}}, \
  and\ \bibinfo {author} {\bibfnamefont {S.}~\bibnamefont {{Das Sarma}}},\
  }\href {\doibase 10.1103/PhysRevLett.105.077001} {\bibfield  {journal}
  {\bibinfo  {journal} {Phys. Rev. Lett.}\ }\textbf {\bibinfo {volume} {105}},\
  \bibinfo {pages} {077001} (\bibinfo {year} {2010})}\BibitemShut {NoStop}%
\bibitem [{\citenamefont {Mourik}\ \emph {et~al.}(2012)\citenamefont {Mourik},
  \citenamefont {Zuo}, \citenamefont {Frolov}, \citenamefont {Plissard},
  \citenamefont {Bakkers},\ and\ \citenamefont {Kouwenhoven}}]{Mourik2012}%
  \BibitemOpen
  \bibfield  {author} {\bibinfo {author} {\bibfnamefont {V.}~\bibnamefont
  {Mourik}}, \bibinfo {author} {\bibfnamefont {K.}~\bibnamefont {Zuo}},
  \bibinfo {author} {\bibfnamefont {S.~M.}\ \bibnamefont {Frolov}}, \bibinfo
  {author} {\bibfnamefont {S.~R.}\ \bibnamefont {Plissard}}, \bibinfo {author}
  {\bibfnamefont {E.~P. A.~M.}\ \bibnamefont {Bakkers}}, \ and\ \bibinfo
  {author} {\bibfnamefont {L.~P.}\ \bibnamefont {Kouwenhoven}},\ }\href
  {\doibase 10.1126/science.1222360} {\bibfield  {journal} {\bibinfo  {journal}
  {Science}\ }\textbf {\bibinfo {volume} {336}},\ \bibinfo {pages} {1003}
  (\bibinfo {year} {2012})}\BibitemShut {NoStop}%
\bibitem [{\citenamefont {Zhang}\ \emph {et~al.}(2018)\citenamefont {Zhang},
  \citenamefont {Liu}, \citenamefont {Gazibegovic}, \citenamefont {Xu},
  \citenamefont {Logan}, \citenamefont {Wang}, \citenamefont {van Loo},
  \citenamefont {Bommer}, \citenamefont {de~Moor}, \citenamefont {Car},
  \citenamefont {{Op het Veld}}, \citenamefont {van Veldhoven}, \citenamefont
  {Koelling}, \citenamefont {Verheijen}, \citenamefont {Pendharkar},
  \citenamefont {Pennachio}, \citenamefont {Shojaei}, \citenamefont {Lee},
  \citenamefont {Palmstr{\o}m}, \citenamefont {Bakkers}, \citenamefont
  {Sarma},\ and\ \citenamefont {Kouwenhoven}}]{Zhang2018}%
  \BibitemOpen
  \bibfield  {author} {\bibinfo {author} {\bibfnamefont {H.}~\bibnamefont
  {Zhang}}, \bibinfo {author} {\bibfnamefont {C.-X.}\ \bibnamefont {Liu}},
  \bibinfo {author} {\bibfnamefont {S.}~\bibnamefont {Gazibegovic}}, \bibinfo
  {author} {\bibfnamefont {D.}~\bibnamefont {Xu}}, \bibinfo {author}
  {\bibfnamefont {J.~A.}\ \bibnamefont {Logan}}, \bibinfo {author}
  {\bibfnamefont {G.}~\bibnamefont {Wang}}, \bibinfo {author} {\bibfnamefont
  {N.}~\bibnamefont {van Loo}}, \bibinfo {author} {\bibfnamefont {J.~D.~S.}\
  \bibnamefont {Bommer}}, \bibinfo {author} {\bibfnamefont {M.~W.~A.}\
  \bibnamefont {de~Moor}}, \bibinfo {author} {\bibfnamefont {D.}~\bibnamefont
  {Car}}, \bibinfo {author} {\bibfnamefont {R.~L.~M.}\ \bibnamefont {{Op het
  Veld}}}, \bibinfo {author} {\bibfnamefont {P.~J.}\ \bibnamefont {van
  Veldhoven}}, \bibinfo {author} {\bibfnamefont {S.}~\bibnamefont {Koelling}},
  \bibinfo {author} {\bibfnamefont {M.~A.}\ \bibnamefont {Verheijen}}, \bibinfo
  {author} {\bibfnamefont {M.}~\bibnamefont {Pendharkar}}, \bibinfo {author}
  {\bibfnamefont {D.~J.}\ \bibnamefont {Pennachio}}, \bibinfo {author}
  {\bibfnamefont {B.}~\bibnamefont {Shojaei}}, \bibinfo {author} {\bibfnamefont
  {J.~S.}\ \bibnamefont {Lee}}, \bibinfo {author} {\bibfnamefont {C.~J.}\
  \bibnamefont {Palmstr{\o}m}}, \bibinfo {author} {\bibfnamefont {E.~P. A.~M.}\
  \bibnamefont {Bakkers}}, \bibinfo {author} {\bibfnamefont {S.~D.}\
  \bibnamefont {Sarma}}, \ and\ \bibinfo {author} {\bibfnamefont {L.~P.}\
  \bibnamefont {Kouwenhoven}},\ }\href {\doibase 10.1038/nature26142}
  {\bibfield  {journal} {\bibinfo  {journal} {Nature}\ }\textbf {\bibinfo
  {volume} {556}},\ \bibinfo {pages} {74} (\bibinfo {year} {2018})}\BibitemShut
  {NoStop}%
\bibitem [{\citenamefont {Wang}\ \emph
  {et~al.}(2018{\natexlab{a}})\citenamefont {Wang}, \citenamefont {Kong},
  \citenamefont {Fan}, \citenamefont {Chen}, \citenamefont {Zhu}, \citenamefont
  {Liu}, \citenamefont {Cao}, \citenamefont {Sun}, \citenamefont {Du},
  \citenamefont {Schneeloch}, \citenamefont {Zhong}, \citenamefont {Gu},
  \citenamefont {Fu}, \citenamefont {Ding},\ and\ \citenamefont
  {Gao}}]{Wang2018b}%
  \BibitemOpen
  \bibfield  {author} {\bibinfo {author} {\bibfnamefont {D.}~\bibnamefont
  {Wang}}, \bibinfo {author} {\bibfnamefont {L.}~\bibnamefont {Kong}}, \bibinfo
  {author} {\bibfnamefont {P.}~\bibnamefont {Fan}}, \bibinfo {author}
  {\bibfnamefont {H.}~\bibnamefont {Chen}}, \bibinfo {author} {\bibfnamefont
  {S.}~\bibnamefont {Zhu}}, \bibinfo {author} {\bibfnamefont {W.}~\bibnamefont
  {Liu}}, \bibinfo {author} {\bibfnamefont {L.}~\bibnamefont {Cao}}, \bibinfo
  {author} {\bibfnamefont {Y.}~\bibnamefont {Sun}}, \bibinfo {author}
  {\bibfnamefont {S.}~\bibnamefont {Du}}, \bibinfo {author} {\bibfnamefont
  {J.}~\bibnamefont {Schneeloch}}, \bibinfo {author} {\bibfnamefont
  {R.}~\bibnamefont {Zhong}}, \bibinfo {author} {\bibfnamefont
  {G.}~\bibnamefont {Gu}}, \bibinfo {author} {\bibfnamefont {L.}~\bibnamefont
  {Fu}}, \bibinfo {author} {\bibfnamefont {H.}~\bibnamefont {Ding}}, \ and\
  \bibinfo {author} {\bibfnamefont {H.-J.}\ \bibnamefont {Gao}},\ }\href
  {\doibase 10.1126/science.aao1797} {\bibfield  {journal} {\bibinfo  {journal}
  {Science}\ }\textbf {\bibinfo {volume} {362}},\ \bibinfo {pages} {333}
  (\bibinfo {year} {2018}{\natexlab{a}})}\BibitemShut {NoStop}%
\bibitem [{\citenamefont {Benalcazar}\ \emph {et~al.}(2017)\citenamefont
  {Benalcazar}, \citenamefont {Bernevig},\ and\ \citenamefont
  {Hughes}}]{Benalcazar2017}%
  \BibitemOpen
  \bibfield  {author} {\bibinfo {author} {\bibfnamefont {W.~A.}\ \bibnamefont
  {Benalcazar}}, \bibinfo {author} {\bibfnamefont {B.~A.}\ \bibnamefont
  {Bernevig}}, \ and\ \bibinfo {author} {\bibfnamefont {T.~L.}\ \bibnamefont
  {Hughes}},\ }\href {\doibase 10.1126/science.aah6442} {\bibfield  {journal}
  {\bibinfo  {journal} {Science}\ }\textbf {\bibinfo {volume} {357}},\ \bibinfo
  {pages} {61} (\bibinfo {year} {2017})}\BibitemShut {NoStop}%
\bibitem [{\citenamefont {Schindler}\ \emph {et~al.}(2018)\citenamefont
  {Schindler}, \citenamefont {Cook}, \citenamefont {Vergniory}, \citenamefont
  {Wang}, \citenamefont {Parkin}, \citenamefont {Bernevig},\ and\ \citenamefont
  {Neupert}}]{Schindler2018}%
  \BibitemOpen
  \bibfield  {author} {\bibinfo {author} {\bibfnamefont {F.}~\bibnamefont
  {Schindler}}, \bibinfo {author} {\bibfnamefont {A.~M.}\ \bibnamefont {Cook}},
  \bibinfo {author} {\bibfnamefont {M.~G.}\ \bibnamefont {Vergniory}}, \bibinfo
  {author} {\bibfnamefont {Z.}~\bibnamefont {Wang}}, \bibinfo {author}
  {\bibfnamefont {S.~S.~P.}\ \bibnamefont {Parkin}}, \bibinfo {author}
  {\bibfnamefont {B.~A.}\ \bibnamefont {Bernevig}}, \ and\ \bibinfo {author}
  {\bibfnamefont {T.}~\bibnamefont {Neupert}},\ }\href {\doibase
  10.1126/sciadv.aat0346} {\bibfield  {journal} {\bibinfo  {journal} {Sci.
  Adv.}\ }\textbf {\bibinfo {volume} {4}},\ \bibinfo {pages} {eaat0346}
  (\bibinfo {year} {2018})},\ \Eprint {http://arxiv.org/abs/1708.03636}
  {1708.03636} \BibitemShut {NoStop}%
\bibitem [{\citenamefont {Langbehn}\ \emph {et~al.}(2017)\citenamefont
  {Langbehn}, \citenamefont {Peng}, \citenamefont {Trifunovic}, \citenamefont
  {von Oppen},\ and\ \citenamefont {Brouwer}}]{Langbehn2017}%
  \BibitemOpen
  \bibfield  {author} {\bibinfo {author} {\bibfnamefont {J.}~\bibnamefont
  {Langbehn}}, \bibinfo {author} {\bibfnamefont {Y.}~\bibnamefont {Peng}},
  \bibinfo {author} {\bibfnamefont {L.}~\bibnamefont {Trifunovic}}, \bibinfo
  {author} {\bibfnamefont {F.}~\bibnamefont {von Oppen}}, \ and\ \bibinfo
  {author} {\bibfnamefont {P.~W.}\ \bibnamefont {Brouwer}},\ }\href {\doibase
  10.1103/PhysRevLett.119.246401} {\bibfield  {journal} {\bibinfo  {journal}
  {Phys. Rev. Lett.}\ }\textbf {\bibinfo {volume} {119}},\ \bibinfo {pages}
  {246401} (\bibinfo {year} {2017})}\BibitemShut {NoStop}%
\bibitem [{\citenamefont {Khalaf}(2018)}]{Khalaf2018}%
  \BibitemOpen
  \bibfield  {author} {\bibinfo {author} {\bibfnamefont {E.}~\bibnamefont
  {Khalaf}},\ }\href {\doibase 10.1103/PhysRevB.97.205136} {\bibfield
  {journal} {\bibinfo  {journal} {Phys. Rev. B}\ }\textbf {\bibinfo {volume}
  {97}},\ \bibinfo {pages} {205136} (\bibinfo {year} {2018})}\BibitemShut
  {NoStop}%
\bibitem [{\citenamefont {Wang}\ \emph
  {et~al.}(2018{\natexlab{b}})\citenamefont {Wang}, \citenamefont {Liu},
  \citenamefont {Lu},\ and\ \citenamefont {Zhang}}]{Wang2018a}%
  \BibitemOpen
  \bibfield  {author} {\bibinfo {author} {\bibfnamefont {Q.}~\bibnamefont
  {Wang}}, \bibinfo {author} {\bibfnamefont {C.-C.}\ \bibnamefont {Liu}},
  \bibinfo {author} {\bibfnamefont {Y.-M.}\ \bibnamefont {Lu}}, \ and\ \bibinfo
  {author} {\bibfnamefont {F.}~\bibnamefont {Zhang}},\ }\href {\doibase
  10.1103/PhysRevLett.121.186801} {\bibfield  {journal} {\bibinfo  {journal}
  {Phys. Rev. Lett.}\ }\textbf {\bibinfo {volume} {121}},\ \bibinfo {pages}
  {186801} (\bibinfo {year} {2018}{\natexlab{b}})}\BibitemShut {NoStop}%
\bibitem [{\citenamefont {Yan}\ \emph {et~al.}(2018)\citenamefont {Yan},
  \citenamefont {Song},\ and\ \citenamefont {Wang}}]{Yan2018}%
  \BibitemOpen
  \bibfield  {author} {\bibinfo {author} {\bibfnamefont {Z.}~\bibnamefont
  {Yan}}, \bibinfo {author} {\bibfnamefont {F.}~\bibnamefont {Song}}, \ and\
  \bibinfo {author} {\bibfnamefont {Z.}~\bibnamefont {Wang}},\ }\href {\doibase
  10.1103/PhysRevLett.121.096803} {\bibfield  {journal} {\bibinfo  {journal}
  {Phys. Rev. Lett.}\ }\textbf {\bibinfo {volume} {121}},\ \bibinfo {pages}
  {096803} (\bibinfo {year} {2018})}\BibitemShut {NoStop}%
\bibitem [{\citenamefont {Shapourian}\ \emph {et~al.}(2018)\citenamefont
  {Shapourian}, \citenamefont {Wang},\ and\ \citenamefont
  {Ryu}}]{Shapourian2018}%
  \BibitemOpen
  \bibfield  {author} {\bibinfo {author} {\bibfnamefont {H.}~\bibnamefont
  {Shapourian}}, \bibinfo {author} {\bibfnamefont {Y.}~\bibnamefont {Wang}}, \
  and\ \bibinfo {author} {\bibfnamefont {S.}~\bibnamefont {Ryu}},\ }\href
  {\doibase 10.1103/PhysRevB.97.094508} {\bibfield  {journal} {\bibinfo
  {journal} {Phys. Rev. B}\ }\textbf {\bibinfo {volume} {97}},\ \bibinfo
  {pages} {094508} (\bibinfo {year} {2018})}\BibitemShut {NoStop}%
\bibitem [{\citenamefont {Liu}\ \emph {et~al.}(2018)\citenamefont {Liu},
  \citenamefont {He},\ and\ \citenamefont {Nori}}]{Liu2018}%
  \BibitemOpen
  \bibfield  {author} {\bibinfo {author} {\bibfnamefont {T.}~\bibnamefont
  {Liu}}, \bibinfo {author} {\bibfnamefont {J.~J.}\ \bibnamefont {He}}, \ and\
  \bibinfo {author} {\bibfnamefont {F.}~\bibnamefont {Nori}},\ }\href {\doibase
  10.1103/PhysRevB.98.245413} {\bibfield  {journal} {\bibinfo  {journal} {Phys.
  Rev. B}\ }\textbf {\bibinfo {volume} {98}},\ \bibinfo {pages} {245413}
  (\bibinfo {year} {2018})}\BibitemShut {NoStop}%
\bibitem [{\citenamefont {Hsu}\ \emph {et~al.}(2018)\citenamefont {Hsu},
  \citenamefont {Stano}, \citenamefont {Klinovaja},\ and\ \citenamefont
  {Loss}}]{Hsu2018}%
  \BibitemOpen
  \bibfield  {author} {\bibinfo {author} {\bibfnamefont {C.-H.}\ \bibnamefont
  {Hsu}}, \bibinfo {author} {\bibfnamefont {P.}~\bibnamefont {Stano}}, \bibinfo
  {author} {\bibfnamefont {J.}~\bibnamefont {Klinovaja}}, \ and\ \bibinfo
  {author} {\bibfnamefont {D.}~\bibnamefont {Loss}},\ }\href {\doibase
  10.1103/PhysRevLett.121.196801} {\bibfield  {journal} {\bibinfo  {journal}
  {Phys. Rev. Lett.}\ }\textbf {\bibinfo {volume} {121}},\ \bibinfo {pages}
  {196801} (\bibinfo {year} {2018})}\BibitemShut {NoStop}%
\bibitem [{\citenamefont {Pan}\ \emph {et~al.}(2019)\citenamefont {Pan},
  \citenamefont {Yang}, \citenamefont {Chen}, \citenamefont {Xu}, \citenamefont
  {Liu},\ and\ \citenamefont {Liu}}]{Pan2019}%
  \BibitemOpen
  \bibfield  {author} {\bibinfo {author} {\bibfnamefont {X.-H.}\ \bibnamefont
  {Pan}}, \bibinfo {author} {\bibfnamefont {K.-J.}\ \bibnamefont {Yang}},
  \bibinfo {author} {\bibfnamefont {L.}~\bibnamefont {Chen}}, \bibinfo {author}
  {\bibfnamefont {G.}~\bibnamefont {Xu}}, \bibinfo {author} {\bibfnamefont
  {C.-X.}\ \bibnamefont {Liu}}, \ and\ \bibinfo {author} {\bibfnamefont
  {X.}~\bibnamefont {Liu}},\ }\href {\doibase 10.1103/PhysRevLett.123.156801}
  {\bibfield  {journal} {\bibinfo  {journal} {Phys. Rev. Lett.}\ }\textbf
  {\bibinfo {volume} {123}},\ \bibinfo {pages} {156801} (\bibinfo {year}
  {2019})}\BibitemShut {NoStop}%
\bibitem [{\citenamefont {Zhang}\ \emph {et~al.}(2019)\citenamefont {Zhang},
  \citenamefont {Cole}, \citenamefont {Wu},\ and\ \citenamefont {{Das
  Sarma}}}]{Zhang2019}%
  \BibitemOpen
  \bibfield  {author} {\bibinfo {author} {\bibfnamefont {R.-X.}\ \bibnamefont
  {Zhang}}, \bibinfo {author} {\bibfnamefont {W.~S.}\ \bibnamefont {Cole}},
  \bibinfo {author} {\bibfnamefont {X.}~\bibnamefont {Wu}}, \ and\ \bibinfo
  {author} {\bibfnamefont {S.}~\bibnamefont {{Das Sarma}}},\ }\href {\doibase
  10.1103/PhysRevLett.123.167001} {\bibfield  {journal} {\bibinfo  {journal}
  {Phys. Rev. Lett.}\ }\textbf {\bibinfo {volume} {123}},\ \bibinfo {pages}
  {167001} (\bibinfo {year} {2019})}\BibitemShut {NoStop}%
\bibitem [{\citenamefont {Wu}\ \emph {et~al.}(2019)\citenamefont {Wu},
  \citenamefont {Liu}, \citenamefont {Thomale},\ and\ \citenamefont
  {Liu}}]{Wu2019}%
  \BibitemOpen
  \bibfield  {author} {\bibinfo {author} {\bibfnamefont {X.}~\bibnamefont
  {Wu}}, \bibinfo {author} {\bibfnamefont {X.}~\bibnamefont {Liu}}, \bibinfo
  {author} {\bibfnamefont {R.}~\bibnamefont {Thomale}}, \ and\ \bibinfo
  {author} {\bibfnamefont {C.-X.}\ \bibnamefont {Liu}},\ }\href
  {http://arxiv.org/abs/1905.10648} {\enquote {\bibinfo {title} {{High-$T_c$
  Superconductor Fe(Se,Te) Monolayer: an Intrinsic, Scalable and
  Electrically-tunable Majorana Platform}},}\ } (\bibinfo {year} {2019}),\
  \Eprint {http://arxiv.org/abs/1905.10648} {arXiv:1905.10648} \BibitemShut
  {NoStop}%
\bibitem [{\citenamefont {Wu}\ \emph {et~al.}(2020)\citenamefont {Wu},
  \citenamefont {Benalcazar}, \citenamefont {Li}, \citenamefont {Thomale},
  \citenamefont {Liu},\ and\ \citenamefont {Hu}}]{Wu2020}%
  \BibitemOpen
  \bibfield  {author} {\bibinfo {author} {\bibfnamefont {X.}~\bibnamefont
  {Wu}}, \bibinfo {author} {\bibfnamefont {W.~A.}\ \bibnamefont {Benalcazar}},
  \bibinfo {author} {\bibfnamefont {Y.}~\bibnamefont {Li}}, \bibinfo {author}
  {\bibfnamefont {R.}~\bibnamefont {Thomale}}, \bibinfo {author} {\bibfnamefont
  {C.-X.}\ \bibnamefont {Liu}}, \ and\ \bibinfo {author} {\bibfnamefont
  {J.}~\bibnamefont {Hu}},\ }\href {https://arxiv.org/pdf/2003.12204.pdf}
  {\enquote {\bibinfo {title} {{Boundary-obstructed topological high-T$_c$
  superconductivity in iron pnictides}},}\ } (\bibinfo {year} {2020}),\ \Eprint
  {http://arxiv.org/abs/2003.12204} {arXiv:2003.12204} \BibitemShut {NoStop}%
\bibitem [{\citenamefont {Hsu}\ \emph {et~al.}(2019)\citenamefont {Hsu},
  \citenamefont {Cole}, \citenamefont {Zhang},\ and\ \citenamefont
  {Sau}}]{Hsu2019}%
  \BibitemOpen
  \bibfield  {author} {\bibinfo {author} {\bibfnamefont {Y.-T.}\ \bibnamefont
  {Hsu}}, \bibinfo {author} {\bibfnamefont {W.~S.}\ \bibnamefont {Cole}},
  \bibinfo {author} {\bibfnamefont {R.-X.}\ \bibnamefont {Zhang}}, \ and\
  \bibinfo {author} {\bibfnamefont {J.~D.}\ \bibnamefont {Sau}},\ }\href
  {http://arxiv.org/abs/1904.06361} {\enquote {\bibinfo {title}
  {{Inversion-protected higher order topological superconductivity in monolayer
  WTe$_2$}},}\ } (\bibinfo {year} {2019}),\ \Eprint
  {http://arxiv.org/abs/1904.06361} {arXiv:1904.06361} \BibitemShut {NoStop}%
\bibitem [{\citenamefont {Khalaf}\ \emph {et~al.}(2018)\citenamefont {Khalaf},
  \citenamefont {Po}, \citenamefont {Vishwanath},\ and\ \citenamefont
  {Watanabe}}]{Khalaf2018a}%
  \BibitemOpen
  \bibfield  {author} {\bibinfo {author} {\bibfnamefont {E.}~\bibnamefont
  {Khalaf}}, \bibinfo {author} {\bibfnamefont {H.~C.}\ \bibnamefont {Po}},
  \bibinfo {author} {\bibfnamefont {A.}~\bibnamefont {Vishwanath}}, \ and\
  \bibinfo {author} {\bibfnamefont {H.}~\bibnamefont {Watanabe}},\ }\href
  {\doibase 10.1103/PhysRevX.8.031070} {\bibfield  {journal} {\bibinfo
  {journal} {Phys. Rev. X}\ }\textbf {\bibinfo {volume} {8}},\ \bibinfo {pages}
  {031070} (\bibinfo {year} {2018})}\BibitemShut {NoStop}%
\bibitem [{\citenamefont {Ono}\ \emph {et~al.}(2019)\citenamefont {Ono},
  \citenamefont {Yanase},\ and\ \citenamefont {Watanabe}}]{Ono2019}%
  \BibitemOpen
  \bibfield  {author} {\bibinfo {author} {\bibfnamefont {S.}~\bibnamefont
  {Ono}}, \bibinfo {author} {\bibfnamefont {Y.}~\bibnamefont {Yanase}}, \ and\
  \bibinfo {author} {\bibfnamefont {H.}~\bibnamefont {Watanabe}},\ }\href
  {\doibase 10.1103/PhysRevResearch.1.013012} {\bibfield  {journal} {\bibinfo
  {journal} {Phys. Rev. Research}\ }\textbf {\bibinfo {volume} {1}},\ \bibinfo
  {pages} {013012} (\bibinfo {year} {2019})}\BibitemShut {NoStop}%
\bibitem [{\citenamefont {Geier}\ \emph {et~al.}(2019)\citenamefont {Geier},
  \citenamefont {Brouwer},\ and\ \citenamefont {Trifunovic}}]{Geier2019}%
  \BibitemOpen
  \bibfield  {author} {\bibinfo {author} {\bibfnamefont {M.}~\bibnamefont
  {Geier}}, \bibinfo {author} {\bibfnamefont {P.~W.}\ \bibnamefont {Brouwer}},
  \ and\ \bibinfo {author} {\bibfnamefont {L.}~\bibnamefont {Trifunovic}},\
  }\href {http://arxiv.org/abs/1910.11271} {\enquote {\bibinfo {title}
  {{Symmetry-based indicators for topological Bogoliubov-de Gennes
  Hamiltonians}},}\ } (\bibinfo {year} {2019}),\ \Eprint
  {http://arxiv.org/abs/1910.11271} {arXiv:1910.11271} \BibitemShut {NoStop}%
\bibitem [{\citenamefont {Shiozaki}(2019)}]{Shiozaki2019}%
  \BibitemOpen
  \bibfield  {author} {\bibinfo {author} {\bibfnamefont {K.}~\bibnamefont
  {Shiozaki}},\ }\href {http://arxiv.org/abs/1907.13632} {\enquote {\bibinfo
  {title} {{Variants of the symmetry-based indicator}},}\ } (\bibinfo {year}
  {2019}),\ \Eprint {http://arxiv.org/abs/1907.13632} {arXiv:1907.13632}
  \BibitemShut {NoStop}%
\bibitem [{\citenamefont {Skurativska}\ \emph {et~al.}(2020)\citenamefont
  {Skurativska}, \citenamefont {Neupert},\ and\ \citenamefont
  {Fischer}}]{Skurativska2020}%
  \BibitemOpen
  \bibfield  {author} {\bibinfo {author} {\bibfnamefont {A.}~\bibnamefont
  {Skurativska}}, \bibinfo {author} {\bibfnamefont {T.}~\bibnamefont
  {Neupert}}, \ and\ \bibinfo {author} {\bibfnamefont {M.~H.}\ \bibnamefont
  {Fischer}},\ }\href {\doibase 10.1103/PhysRevResearch.2.013064} {\bibfield
  {journal} {\bibinfo  {journal} {Phys. Rev. Research}\ }\textbf {\bibinfo
  {volume} {2}},\ \bibinfo {pages} {013064} (\bibinfo {year}
  {2020})}\BibitemShut {NoStop}%
\bibitem [{\citenamefont {Ono}\ \emph {et~al.}(2020)\citenamefont {Ono},
  \citenamefont {Po},\ and\ \citenamefont {Watanabe}}]{Ono2020}%
  \BibitemOpen
  \bibfield  {author} {\bibinfo {author} {\bibfnamefont {S.}~\bibnamefont
  {Ono}}, \bibinfo {author} {\bibfnamefont {H.~C.}\ \bibnamefont {Po}}, \ and\
  \bibinfo {author} {\bibfnamefont {H.}~\bibnamefont {Watanabe}},\ }\href
  {\doibase 10.1126/sciadv.aaz8367} {\bibfield  {journal} {\bibinfo  {journal}
  {Sci. Adv.}\ }\textbf {\bibinfo {volume} {6}},\ \bibinfo {pages} {eaaz8367}
  (\bibinfo {year} {2020})}\BibitemShut {NoStop}%
\bibitem [{\citenamefont {Zhang}\ \emph {et~al.}(2020)\citenamefont {Zhang},
  \citenamefont {Sau},\ and\ \citenamefont {{Das Sarma}}}]{Zhang2020}%
  \BibitemOpen
  \bibfield  {author} {\bibinfo {author} {\bibfnamefont {R.-X.}\ \bibnamefont
  {Zhang}}, \bibinfo {author} {\bibfnamefont {J.~D.}\ \bibnamefont {Sau}}, \
  and\ \bibinfo {author} {\bibfnamefont {S.}~\bibnamefont {{Das Sarma}}},\
  }\href {http://arxiv.org/abs/2003.02559} {\enquote {\bibinfo {title} {{Kitaev
  Building-block Construction for Higher-order Topological Superconductors}},}\
  } (\bibinfo {year} {2020}),\ \Eprint {http://arxiv.org/abs/2003.02559}
  {arXiv:2003.02559} \BibitemShut {NoStop}%
\bibitem [{\citenamefont {Zhang}\ \emph {et~al.}(2013)\citenamefont {Zhang},
  \citenamefont {Kane},\ and\ \citenamefont {Mele}}]{Zhang2013}%
  \BibitemOpen
  \bibfield  {author} {\bibinfo {author} {\bibfnamefont {F.}~\bibnamefont
  {Zhang}}, \bibinfo {author} {\bibfnamefont {C.~L.}\ \bibnamefont {Kane}}, \
  and\ \bibinfo {author} {\bibfnamefont {E.~J.}\ \bibnamefont {Mele}},\ }\href
  {\doibase 10.1103/PhysRevLett.111.056402} {\bibfield  {journal} {\bibinfo
  {journal} {Phys. Rev. Lett.}\ }\textbf {\bibinfo {volume} {111}},\ \bibinfo
  {pages} {056402} (\bibinfo {year} {2013})}\BibitemShut {NoStop}%
\bibitem [{\citenamefont {Schnyder}\ and\ \citenamefont
  {Ryu}(2011)}]{Schnyder2011}%
  \BibitemOpen
  \bibfield  {author} {\bibinfo {author} {\bibfnamefont {A.~P.}\ \bibnamefont
  {Schnyder}}\ and\ \bibinfo {author} {\bibfnamefont {S.}~\bibnamefont {Ryu}},\
  }\href {\doibase 10.1103/PhysRevB.84.060504} {\bibfield  {journal} {\bibinfo
  {journal} {Phys. Rev. B}\ }\textbf {\bibinfo {volume} {84}},\ \bibinfo
  {pages} {060504} (\bibinfo {year} {2011})}\BibitemShut {NoStop}%
\bibitem [{\citenamefont {Wong}\ \emph {et~al.}(2013)\citenamefont {Wong},
  \citenamefont {Liu}, \citenamefont {Law},\ and\ \citenamefont
  {Lee}}]{Wong2013}%
  \BibitemOpen
  \bibfield  {author} {\bibinfo {author} {\bibfnamefont {C.~L.~M.}\
  \bibnamefont {Wong}}, \bibinfo {author} {\bibfnamefont {J.}~\bibnamefont
  {Liu}}, \bibinfo {author} {\bibfnamefont {K.~T.}\ \bibnamefont {Law}}, \ and\
  \bibinfo {author} {\bibfnamefont {P.~A.}\ \bibnamefont {Lee}},\ }\href
  {\doibase 10.1103/PhysRevB.88.060504} {\bibfield  {journal} {\bibinfo
  {journal} {Phys. Rev. B}\ }\textbf {\bibinfo {volume} {88}},\ \bibinfo
  {pages} {060504} (\bibinfo {year} {2013})}\BibitemShut {NoStop}%
\end{thebibliography}%
	
	\onecolumngrid
	\appendix
    \section{Majorana Bonding and Wannier orbitals}
	In this section, we discuss the relation between Majorana bonding and the position of resulting Wannier orbital pair. We assume two Majorana Krammer pairs ($\alpha_{s},\alpha_{-s}$) and ($\beta_{s'},\beta_{-s'}$) localized at $\mathbf{R}_\alpha$ and $\mathbf{R}_\beta$ respectively. The interaction between the two pairs in the basis ($\alpha_{s},\alpha_{-s},\beta_{s'},\beta_{-s'}$) is 
	\begin{equation}
	T=\begin{pmatrix}
	0 & A\\ A^\dagger &0
	\end{pmatrix}
	\end{equation}
	where $A=-A^\ast$ due to the anti-commutation of Majorana fermions. $T$ has vanishing diagonal matrix elements because time-reversal invariance forbids coupling between time-reversed partners. The Hamiltonian $T$ thus has four solutions, corresponding to two Dirac fermions
	\begin{equation}
	\nu_1=\begin{pmatrix}
	u_1\\ v_1
	\end{pmatrix}, \quad \nu_2=\begin{pmatrix}
	u_1^\ast\\ v_1^\ast
	\end{pmatrix}, \quad \nu_3=\begin{pmatrix}
	u_2\\ v_2
	\end{pmatrix}, \quad \nu_4=\begin{pmatrix}
	u_2^\ast\\ v_2^\ast
	\end{pmatrix},
	\end{equation}
	where $u_1,u_2,v_1,v_2$ are two-component vectors. The position relative to the bond center is given by the matrix $X=\mathbf{R} \sigma_0\otimes\sigma_z$ where $\mathbf{R}=(\mathbf{R}_\alpha-\mathbf{R}_\beta)/2$. From the eigenvector equation, we have
	\begin{equation}
	\begin{cases}
	Av = Eu\\
	A^\dagger u = E v
	\end{cases} \Rightarrow |u|^2=|v^2| \Rightarrow u = \mathcal{U}v,
	\end{equation}
	with $\mathcal{U}$ being unitary. As a result, we can define a transformation
	\begin{equation}
	M=\begin{pmatrix}
	0 & \mathcal{U}^\dagger \\ \mathcal{U} & 0
	\end{pmatrix},
	\end{equation}
	so that $M \nu_1 = \nu_1$ but $\{M,X\}=0$. Thus $(\nu_1^\ast)^TX \nu_1=0$ and similarly,  $(\nu_3^\ast)^TX \nu_3=0$. As a result, we have shown that the Wannier orbital pair is always localized at the middle of the bond, regardless of the bond type.

    \section{Analytic phase diagram of the model}
	
	The time-reversal-invariant 2D Hamiltonian with $l=A/B$ as the sub-lattice index is given by
	\begin{equation}
	\begin{split}
	H=&i\sum_{\mathbf{R},l} \beta^l_{\mathbf{R},s}\left(t_{0\parallel}^l\sigma_0+it_{0\times}^l\sigma_y\right)_{ss'}\alpha^l_{\mathbf{R},s'} + \beta^l_{\mathbf{R},s}\left(t_{x\parallel}^l\sigma_0+it_{x\times}^l\sigma_y\right)_{ss'} \alpha^l_{\mathbf{R}+\mathbf{a},s'}\\
	\quad &+i\sum_{\mathbf{R}} \beta_{\mathbf{R},s}^A\left(t_{y\parallel}\sigma_z+t_{y\times}\sigma_x\right)_{ss'}\beta_{\mathbf{R},s'}^B - \alpha_{\mathbf{R},s}^B\left( t_{y\parallel}\sigma_z-t_{y\times}\sigma_x\right)\alpha_{\mathbf{R}+\mathbf{b},s'}^A.
	\end{split}
	\end{equation}
	Without loss of generality, we take $t_{0\parallel}^A=t_{0\parallel}^B=t_{0\parallel}$, $t_{0\times}^A=-t_{0\times}^B=t_{0\times}$, $t_{x\parallel}^A=t_{x\parallel}^B=t_{x\parallel}$, and $t_{x\times}^A=-t_{x\times}^B=-t_{x\times}$. We can then define a local transformation to decouple the spinful system into two time-reversal-related copies:
	\begin{equation}
	\begin{split}
	& \ket{\alpha_{m,n}^A} = O[(\theta_0-\theta_x)m-\theta_0/2+\theta_y(2n-1)] \ket{\alpha'^A_{m,n}}, \\
	& \ket{\alpha_{m,n}^B} = O[-(\theta_0-\theta_x)m+\theta_0/2-2\theta_yn] \ket{\alpha'^B_{m,n}}, \\
	& \ket{\beta_{m,n}^A} = O[(\theta_0-\theta_x)m+\theta_0/2+\theta_y(2n-1)] \ket{\beta'^A_{m,n}}, \\
	& \ket{\beta_{m,n}^B} = O[-(\theta_0-\theta_x)m-\theta_0/2-2\theta_yn] \ket{\beta'^B_{m,n}}; \\   
	\end{split}
	\end{equation}
	where $O[\theta]=\cos\theta \sigma_0 +i\sin\theta\sigma_y$ is the rotation matrix, $\ket{\alpha}$ and $\ket{\beta}$ denote the two-component spinors for brevity, and $\theta_0=\text{Arg}(t_{x\parallel}+it_{x\times})$, $ \theta_x=\text{Arg}(t_{x\parallel}+it_{x\times})$, $\theta_y=\text{Arg}(t_{y\parallel}+it_{y\times})$. Under such a transformation, the Hamiltonian becomes a direct sum of two effectively ``spinless" Hamiltonian $H=h\bigoplus h$ where
	\begin{equation}
	\begin{split}
	h=&i\sum_{m,n,l}\left(t_0\beta'^l_{m,n}\alpha'^l_{m,n} + t_x\beta'^l_{m,n}\alpha'^l_{m+1,n}\right) +it_y\sum_{m,n}\left( \beta'^A_{m,n}\beta'^B_{m,n} - \alpha'^B_{m,n}\alpha'^A_{m,n+1}\right),
	\end{split}
	\end{equation}
	and $t_0=\sqrt{t_{0\parallel}^2+t_{0\times}^2}$, $t_x=\sqrt{t_{x\parallel}^2+t_{x\times}^2}$, $t_y=\sqrt{t_{y\parallel}^2+t_{y\times}^2}$. The bulk gap closing of $h$ only depends on three parameters $t_{0,x,y}$, instead of the previous six parameters in the original Hamiltonian. In particular, the system becomes nodal when $|t_0-t_x|\le t_y \le t_0+t_x$.
	
	\section{Nodal superconductor}
		When $|t_0-t_x|<t_y<t_0+t_x$, the system hosts four nodal points at
	\begin{equation}
	\begin{cases}
	k_y= 2\theta_y, k_x=\pm k_0 + (\theta_0-\theta_x)\\
	k_y=-2\theta_y, k_x=\pm k_0 + (\theta_x-\theta_0)
	\end{cases},
	\end{equation}
	where $k_0=\cos^{-1}\left[(t_y^2-t_0^2-t_x^2)/(2t_0t_x)\right]$. The $\mathcal{P}$ and $\Theta$ naturally define a chiral symmetry $\mathcal{C}=i\tau_x\otimes\sigma_y\otimes\gamma_0$. Then
	\begin{equation}
	\begin{split}
	&\mathcal{C}=U_C\mathcal{D}_CU_C^\dagger,\quad U_C^\dagger H(k)U_C = \begin{pmatrix}
	0 & \mathcal{N}(k) \\
	\mathcal{N}^\dagger(k) & 0
	\end{pmatrix}\\
	&\mathcal{N}(k) = U(k)\mathcal{D}V(k)^{\dagger}, \quad A(k)= U(k)V(k)^{\dagger}
	\end{split}
	\end{equation}
	where $\mathcal{D}$ denotes a diagonal matrix. The winding number around a Dirac point is given by \cite{Schnyder2011}
	\begin{equation}
	\nu = \frac{1}{2\pi}\oint \nabla_{\mathbf{k}}\text{Arg}\{\det[A(\mathbf{k})]\} d\mathbf{k}.
	\end{equation}
	When $\theta_x\ne\theta_0$ and $\theta_y\ne 0$ and $\pi/2$, there are four separate Dirac points with non-zero winding numbers shown in the main text. Due to the anti-commutation $\{i\sigma_y,U_C \}=0$, we have
	$\mathcal{N}^\text{T}(-\mathbf{k}) =  \mathcal{T}^\dagger \mathcal{N}(\mathbf{k}) \mathcal{T}$,
	with $\mathcal{T}$ being a unitary matrix. As a result, $\det [A(\mathbf{k})] = \det [A(-\mathbf{k})]$ and the winding numbers around two opposite-momentum Dirac points are the same.

\end{document}